\magnification\magstep1
\overfullrule = 0 pt

\centerline{\bf  ALL CONFORMALLY FLAT   PURE RADIATION METRICS.
}

\

\centerline{ S. Brian Edgar\footnote{$^1$}{ Department of Mathematics,
University of Link\"oping, Link\"oping, Sweden S581 83}
and
Garry Ludwig\footnote{$^2$}
{ Department of Mathematical Sciences, University of Alberta,
Edmonton, Alberta, Canada  T6G 2G1}
 }

\

\

\baselineskip20pt
\parindent=0pt
\centerline{\bf Abstract}\parindent=0pt

\medskip

The complete class of conformally flat, pure radiation metrics is given,
generalising the metric recently given by Wils.

\vfill\eject

 Wils [1] has recently given an
example of a conformally flat metric representing pure radiation (i.e.
energy-momentum tensor $T_{ij}=\Phi^2 l_il_j$, where $l_i$ is a null vector),
 and he
 points out, since this metric is not a plane wave, it
contradicts  Theorem 32.17 in Kramer et al. [2]. In the  coordinate system
$(u,v,x,y)$ this metric is given by
$$g_{ij}=\pmatrix{-2f(u) x(x^2+y^2 )+{v^2/ x^2}  &{  -1} & 2{v/ x}&
0\cr
 -1 & 0 & 0 & 0\cr 2{v/ x} & 0&1&0\cr
0&0&0&1}
\eqno(1)$$
where $f(u)$ is an arbitrary function of $u$, which cannot be zero in curved
spacetime.

Using a coordinate transformation $w=v/x$, Koutras and McIntosh [3] (see
also [4]) have recently written this metric in a slightly different
coordinate system
$(u,w,x,y)$,
$$g_{ij}=\pmatrix{-2f(u) x(x^2+y^2)+w^2 &{ -x}
&w & 0\cr  -x & 0 & 0 & 0\cr
w & 0 &1&0\cr
0&0&0&1}
\eqno(2)
$$
They have also presented what appears to be a
generalisation of Wils's metric, which is also a conformally flat metric
representing pure radiation,
$$g_{ij}=\pmatrix{-2f(u) (ax+b)(x^2+y^2)+a^2w^2 &{ -ax-b}
&aw & 0\cr  -ax-b & 0 & 0 & 0\cr
aw & 0 &1&0\cr
0&0&0&1}
\eqno(3)
$$
where $a,b$ are parameters.
This would seem to contradict Wils's claim that his
example completes the class of    conformally flat  pure radiation metrics,
which are not plane waves.

Moreover, it is easy to confirm (eg. using Maple) that what appears to be an
even more general metric,
$$g_{ij}=\pmatrix{A(u,w,x,y) &{ -a(u)x-b(u)}
&a(u)w & 0\cr   -a(u)x-b(u) & 0 & 0 & 0\cr
a(u)w & 0 &1&0\cr
0&0&0&1}
\eqno(4)$$
where
$$A (u,w,x,y) = -2f(u)\Bigl( a(u) x+ b(u)\Bigr)\Bigl(x^2+y^2 +k(u)x+g(u) y
+l(u) w +h(u)\Bigr)+a^2(u) w^2
\eqno(5)$$
and $a,b,f,g,k,l$  are arbitrary
functions of $u$, is also a conformally flat metric representing pure radiation;
but a little manipulation shows that,
by a coordinate transformation  of the type,
$$ w\rightarrow \alpha(u)w+\beta(u)x+\gamma(u),
\qquad u\rightarrow \lambda(u),\qquad x\rightarrow x+{b(u)\over a(u)}
\eqno(6)$$
this metric can be  transformed into the simpler form,
$$g_{ij}=\pmatrix{-2f(u) x\Bigl(x^2+y^2 +g(u) y
+h(u)\Bigr)+  w^2 &{ -x}
& w & 0\cr   -x & 0 & 0 & 0\cr
w & 0 &1&0\cr
0&0&0&1}
\eqno(7)$$
or, equivalently, in $u,v,x,y$ coordinates,

$$g_{ij}=\pmatrix{-2f(u) x\bigl(x^2+y^2 +g(u) y
+h(u)\bigr)+{v^2/ x^2}  &{  -1} & {2v/
x}& 0\cr
 -1 & 0 & 0 & 0\cr {2v/ x} & 0&1&0\cr
0&0&0&1}
\eqno(8)$$
where
$f(u),g(u),h(u)$ are arbitrary functions of $u$.

\medskip

We believe this metric (8) is a non-trivial generalisation of Wils
conformally flat radiation metric, and indeed represents the whole class of
conformally flat radiation metrics which are not plane waves.
In curved space
$f(u)\ne0$, although  $g(u), h(u)$ can be zero; clearly when  both are
zero (8) reduces to the Wils metric given by (1).

It is a little surprising
that the metric (8) has been overlooked by Kramer et al. in Chapter 32, [2],
since they seem to come close to giving it explicitly in their earlier
Sections 27.5,6, as we shall now show.

In Chapter 27 Kramer et al. [2] discuss the Kundt class [5] of spacetimes,
i.e.  spacetimes admitting a null
congruence
$l_i$ which is divergence-free ($\rho=0$), and which
therefore has a metric of the form
$$ds^2= 2 d\zeta d\bar \zeta - 2du(dv + W d \zeta + \bar W d \bar \zeta + H du)
\eqno(9)$$
where we are using the notation of Chapter 27 in [2]. For the vacuum, pure
radiation,
or Einstein-Maxwell subclasses
the Weyl tensor  is algebraically special;  pure radiation or
null Einstein-Maxwell fields are aligned.

Within the Kundt class, for vacuum, pure radiation,
or null Einstein-Maxwell fields, when the  Weyl tensor is
specialised to either Petrov type III, N or O, the field   equations then lead
to the two separate cases, [2]:

(i) $$W= W^o(\bar \zeta,u), \qquad H = H^o(\zeta,\bar \zeta,u) +(W_{,\bar
\zeta}+\bar W_{,
\zeta})v
\eqno(10)$$
with the non-zero Weyl tensor components
$$\Psi_3={1\over 2}W_{,\bar \zeta \bar \zeta}, \qquad\qquad \Psi_4=H_{,\bar
\zeta
\bar
\zeta}
\eqno(11)$$
It is shown in [2] that, for Petrov type N or O, we are able to use the
remaining coordinate-tetrad freedom to put
$W=0$. Therefore, we can conclude that the only solutions of Petrov type O
are a subset of the plane-fronted waves, subject to the  condition
$$ H^o_{,\bar \zeta  \bar \zeta}=0
\eqno(12)$$

(ii)  $$W= W^o(\bar \zeta,u) - {2v \over
\zeta +
\bar
\zeta}{} \ ,
\qquad H = H^o(\zeta,\bar \zeta,u) +{{W^o+\bar W^o}\over \zeta + \bar
\zeta}v- {v^2
\over (\zeta +
\bar \zeta)^2}
\eqno(13)$$
with the  non-zero Weyl tensor components
$$\Psi_3={\bar W^o_{,\bar \zeta }\over (\zeta + \bar \zeta)} , \quad\qquad
\Psi_4= (\zeta+\bar\zeta)\Bigl({ H^o\over
\zeta+\bar\zeta}\Bigr)_{,\bar \zeta
\bar
\zeta}
\eqno(14)$$
It is shown in [2] that, for Petrov type N or O, we are able to
use some of the remaining coordinate-tetrad freedom to put $W^o=0$.
Therefore, we can conclude that the only solutions of Petrov type O must
satisfy the  condition
$$ \Bigl({ H^o\over
\zeta+\bar\zeta}\Bigr)_{,\bar \zeta
\bar
\zeta}=0
\eqno(15)$$
We can
easily integrate this to obtain
$$ H^o=
f(u)x\Bigl(x^2+y^2 +2\sqrt 2\Re( Y(u))x- 2\sqrt 2\Im( Y(u))y+ h(u)\Bigr)
\eqno(16)$$
giving
$$ H=
f(u)x\Bigl(x^2+y^2 +2\sqrt 2\Re( Y(u))x- 2\sqrt 2\Im( Y(u))y+ h(u)\Bigr)
-{v^2
\over 2x^2}
\eqno(17)$$
where $Y$ is an arbitrary complex function of $u$,  $h$ is an arbitrary real
function of
$u$, and we have substituted $\zeta=(x+iy)/\sqrt 2$.
 The remaining coordinate freedom (see equation (27.11b) in [2]) can be
used to put
$\Re (Y(u))=0 $, and
the metric in (9) then becomes (8) under the
substitution (17), with $ g(u)=-2\sqrt 2\Im(Y(u))$.

Therefore the metric (8) represents all conformally flat  pure
radiation metrics
in the Kundt class, which are not plane waves.
Furthermore, it follows
immediately from the Bianchi equations that all conformally flat  pure
radiation metrics are divergence-free, and thus fall into Kundt's
class, [5]. So  the class of metrics (8) --- together with the class of plane
waves given above by (9), subject to the substitution $W=0$ and $H=H^o$,
where
$H^o$ is obtained from (12)
---
 completes the class of
conformally flat metrics representing pure
radiation.

\medskip
For completeness, we add the following comments:

\medskip
(1)  Einstein-Maxwell null fields, which are a subset of the pure
radiation fields
obtained by applying Maxwell's equations, do not exist for the class of metrics
(8); nor does this class permit massless scalar fields, nor neutrino fields.

\medskip
(2) It is easy to deduce from the above discussion (and is implicit in Chapter
25, [2]) that the
 class of divergence-free pure radiation metrics whose Weyl tensor is
of Petrov  type  N is also obtained from (9) in two separate cases as follows,

(i) $W=0$
and
$H(u,\zeta,\bar \zeta)$ satisfying $H_{,\bar \zeta  \bar \zeta}\ne 0$. This
is clearly a class of plane waves.

(ii)
$$g_{ij}=\pmatrix{-H^o(u,\zeta,\bar \zeta)+{v^2\over x^2}  &{  -1} & {2v\over
x}& 0\cr
 -1 & 0 & 0 & 0\cr {2v\over x} & 0&1&0\cr
0&0&0&1}
\eqno(18)$$
and
$H^o(u,\zeta,\bar \zeta)$ satisfying
$\Bigl({H^o\over \bar \zeta + \zeta}\Bigr)_
{,\bar \zeta  \bar \zeta}\ne 0$.

Once again, it follows
immediately from the Bianchi equations that the Type N  pure
radiation metrics are divergence-free, and thus fall into Kundt's class.
So  the two metrics just given are the complete  class of
Petrov type N  pure
radiation metrics.

\medskip
(3) The case of Petrov type III spacetimes has been discussed
alongside Petrov II in Chapter 25, [2] and determined up to a degree of
coordinate freedom. However, this is not necessarily the complete class of
Petrov
type III pure radiation metrics, but rather only the  class of Petrov type III
pure radiation metrics in the Kundt class, i.e. whose ray
vector is divergence-free ($\rho=0$);
Petrov type III pure radiation metrics --- unlike Petrov O and N --- do not
necessarily have a divergence-free ray
vector.

\

Of course, there is the possibility that this new metric (8) is simply
Wils' metric in a different coordinate system. However, we note in Wils'
derivation --- which was for a more general situation of a conformally Ricci
flat type N pure radiation space --- that he obtained a condition
for $H^o$ (equation (3.8) in [1]) which is supposed to be the condition for
{\it conformally Ricci flat} metrics. Clearly this condition should be
satisfied automatically for {\it conformally flat} metrics; however, our
solution (16) for $H^o$ does not satisfy this condition, in general.
Therefore, since our expression  for $H^o$ given by (16) leads to a
 conformally flat metric, but fails to  satisfy equation (3.8)
in [1] in general,
Wils' condition (3.8)
 must be unnecessarily strong. Further, since Wils' metric in [1] satisfies
(3.8)  it must be less general than ours.

\

In addition,  Skea [7] has used the invariant classification of the
CLASSI program [8] to prove that the space-time (8) only reduces to the  Wils
case (1) when the functions g(u) and h(u) are constants. The detailed
classification by Skea of this metric, and of Wils metric [1],  is available at
the on-line exact solutions database in Brazil and in Canada, [7].

\

In a subsequent paper we shall present, via the GHP formalism, an
alternative simple derivation of the metric found here and discuss in detail
how to extend Wils' results.

\

\

{\bf Acknowledgements.}

One author (BE) would like to thank the  Department of Mathematical Sciences,
University of Alberta for its hospitality while part of this work was
being carried out, and for travel support from the Swedish Natural
Science Research Council. The other author (GL) is grateful for the
continuing financial support by the Natural Sciences and Engineering
Research Council of Canada.

\

{\bf References.}

1. Wils P. (1989) {\it Class. Quantum Grav.} {\bf 6,} 1243.

2. Kramer D., Stephani H., MacCallum M. and Herlt E. (1980) {\it Exact
Solutions of Einstein's Equations} (Cambridge: Cambridge University Press).

3. Koutras A. and McIntosh C.  (1996) {\it Class. Quantum Grav.} {\bf 13,}
L47.

4. Koutras A. (1992) {\it Class. Quantum Grav.} {\bf 9,} L143.

5. Kundt W. (1962) {\it Proc. Roy. Soc. Lond. } {\bf A 270,} 328.

6. Geroch R., Held A., and Penrose R. (1973). {\it J. Math. Phys.,} {\bf
14,} 874.

7. Skea J. {\it private communication}, and at http://edradour.symbcomp.uerj.br
and

http://www.astro.queensu.ca/\~{}jimsk.

8. \AA man J.E. (1982).  {\it Manual for  CLASSI: Classification program for
geometries in General Relativity.} Preprint,  University of Stockholm.

\

\end